\documentclass{emulateapj}
\usepackage{amssymb}
\usepackage{amsmath}

\usepackage[colorinlistoftodos]{todonotes}
\usepackage{xspace}

\newcommand{\be}{\begin{equation}}
\newcommand{\ee}{\end{equation}}

\newcommand{\jybm}{Jy\,beam$^{-1}$\xspace}
\newcommand{\kms}{km\,s$^{-1}$\xspace}
\newcommand{\jybmkms}{Jy\,beam$^{-1}$\,km\,s$^{-1}$\xspace}

\shorttitle{ALMA observations of HD\,100546}
\shortauthors{J.E.Pineda et al.}

\begin{document}

\title{Resolved images of the protoplanetary disk around HD\,100546 with ALMA}

\author{Jaime E. Pineda\altaffilmark{1}, 
Sascha P.\,Quanz\altaffilmark{1}, 
Farzana Meru\altaffilmark{1},  
Gijs D. Mulders\altaffilmark{2}, 
Michael R. Meyer\altaffilmark{1}, 
Olja Pani{\'c} \altaffilmark{3},
\and
Henning Avenhaus\altaffilmark{1}
}
\altaffiltext{1}{Institute for Astronomy, ETH Zurich, Wolfgang-Pauli-Strasse 27, 8093 Zurich, Switzerland}
\email{pjaime@phys.ethz.ch} 
\altaffiltext{2}{Lunar and Planetary Laboratory, The University of Arizona, 1629 E. University Blvd., Tucson, AZ 85721, USA}
\altaffiltext{3}{Institute of Astronomy, Madingley Road, Cambridge, CB3 0HA, UK}

\begin{abstract}
The disk around the Herbig Ae/Be star HD\,100546 has been extensively studied and it is one of the 
systems for which there are observational indications of ongoing and/or recent planet formation.
However, up until now no resolved image of the millimeter dust emission or the gas has been published. 
We present the first resolved images of the disk around HD\,100546 obtained in Band 7 with the ALMA observatory. 
The CO (3--2) image reveals a gas disk that extends out to 350\,au radius at the 3-$\sigma$ level. 
Surprisingly, the 870\,$\mu$m dust continuum emission is compact (radius $<$60\,au) and asymmetric. 
The dust emission is well matched by a truncated disk with outer radius of $\approx$50\,au. 
The lack of millimeter-sized particles outside the 60\,au is consistent with radial drift of particles of this size. 
The protoplanet candidate, identified in previous high-contrast NACO/VLT $L^\prime$ observations, 
could be related to the sharp outer edge of the millimeter-sized particles. 
Future higher angular resolution ALMA observations are needed to determine the detailed properties 
of the millimeter emission and the gas kinematics in the inner region ($<$2$\arcsec$). 
Such observations could also reveal the presence of a planet through the detection of circumplanetary disk material.
\end{abstract}


\section{Introduction} \label{Intro}
Gas and dust rich disks around young stars are the birthplace of new planetary systems. 
However, we still lack observational data showing under which physical and chemical conditions planet formation takes place.

One object where this might be possible is the Herbig Ae/Be star HD\,100546, 
which is located at a distance of 97$\pm$4\,pc \citep{vanLeeuwen_2007-Hipparcos_update}. 
The transition disk around this star has a cavity (in dust and molecular gas) between $\sim$1--14\,au \citep[e.g.,][]{bouwman2003,grady2005,benisty2010,quanz2011,Mulders_2013-HD100546_Companion_Mass,Panic_2014-HD100546_Disk_Asymmetry,liskowsky2012,brittain2009,vanderplas2009,Liu_2003-HD100546_Resolved_Disk}.Various studies suggested the presence of a companion inside this cavity \citep[e.g.,][]{bouwman2003,Acke_2006-VLT_HD100546_Rotation_Companion,tatulli2011,brittain2013,Mulders_2013-HD100546_Companion_Mass}.
Recently, \cite{Quanz_2013-NACO_HD100546} revealed an additional protoplanet candidate further out 
in the outer disk  ($\sim$50--60\,au separation from the central star) using high-contrast direct imaging observations.

To determine the main, large-scale kinematic properties of the disk, \cite{Acke_2006-VLT_HD100546_Rotation_Companion} used 
echelle spectra at optical wavelengths. 
They determined that the major axis is located at 150$\pm$11$\degr$  east of north. 
Other observations of HD\,100546 using single dish millimeter data to study the large scale gas kinematics 
\citep{Panic_2010-HD100546_Kinematics,Panic_2009-Disk_Models} 
show that the observed asymmetric profiles could arise from a warped disk,  
where one side is colder due to a shadow casted by the inner rim.
Herschel/PACS observations of high-J CO lines also requier a two temperature component model 
to reproduce the CO emission \citep{Meeus_2013-DIGIT_CO}.

The current best dust continuum data published for HD\,100546 are from 
\cite{Henning_1998-Survey_HAeBe} and \cite{Wilner_2003-ATCA_TWHydra_HD100546}.
\cite{Henning_1998-Survey_HAeBe} observed HD\,100546 at 1.3\,mm and 
detected the disk millimeter emission (69\,mJy) with 23$\arcsec$ resolution. 
\cite{Wilner_2003-ATCA_TWHydra_HD100546} observed HD\,100546 using the Australia Telescope Compact Array (ATCA) 
at 3.4\,mm and with an angular resolution of 2$\arcsec$. 
They detected the dust continuum emission (36$\pm$3\,mJy) but could not determine the morphology.

While the disk around HD\,100546 has been successfully resolved between $\sim$10 and $\sim$350\,au in scattered light at optical and near-infrared , 
as well as in thermal infrared continuum\footnote{Some studies detected scattered light even at much larger radii, but it seems more likely that they revealed the remainings of a large-scale envelope rather than a genuine circumstellar disk.} 
\citep[e.g.,][Avenhaus et al., submitted]{pantin2000,augereau2001,grady2001,Liu_2003-HD100546_Resolved_Disk,Leinert_2004-Disks_MIDI,ardila2007,quanz2011,Mulders_2011-HD100546_Model,Boccaletti_2013-HD100546_Gemini_Multiple_Spiral}, 
so far no spatially resolved images of the dust and gas content at (sub)-mm wavelengths 
have been published. 
Here we present an analysis of archival ALMA Cycle 0 observations of the 
870\,$\mu$m dust continuum and CO(3--2) emission. 
These data and a comparison with previously published scattered light images  
provide a better determination of fundamental disk properties and hence allow us to 
better constrain the environment in a likely planet forming disk.
A more detailed analysis of these data with a focus on the disk chemistry can be found in Walsh et al. (in prep).

\section{Data}
HD\,100546 was observed on November 18, 2012 with ALMA using Band 7 receivers under 
project 2011.0.00863.S. 
The array configuration included 24 antennas, with baselines between 21 and 375--m. 
The observations cycled through HD\,100546 and quasar J1147$-$6753, with a cycle time of $\sim$10 minutes. 
The bright quasar 3C~279 is used as bandpass calibrator, while Titan is used to set the flux amplitude. 
The standard flagging and calibration was done using CASA 3.4 \citep{McMullin_2007-CASA}, 
while imaging was done using CASA 4.2. 
The 870\,$\mu$m continuum was obtained from line free channels and 
imaged using uniform weighting to achieve an 
angular resolution of 0.93$\arcsec \times$0.37$\arcsec$ (PA=39$\degr$). 
The rms noise is 5.7\,m\jybm, as estimated from emission free regions.

The CO~(3--2) data cube is obtained from the continuum subtracted visibilities 
obtained using the task \verb+uvcontsub+. 
The imaging was done using multiscale clean with uniform weighting, which produced 
a beam size of 0.92$\arcsec \times$0.38$\arcsec$ (PA=38$\degr$). 
In the spectral cube the rms, obtained from line-free channels,  is 27\,m\jybm\,per channel, 
with a channel width of 0.1058\,\kms and a spectral resolution of two channels.

\section{Results} \label{Results}

\subsection{Dust}
\begin{figure*}[ht]
\includegraphics[width=\textwidth]{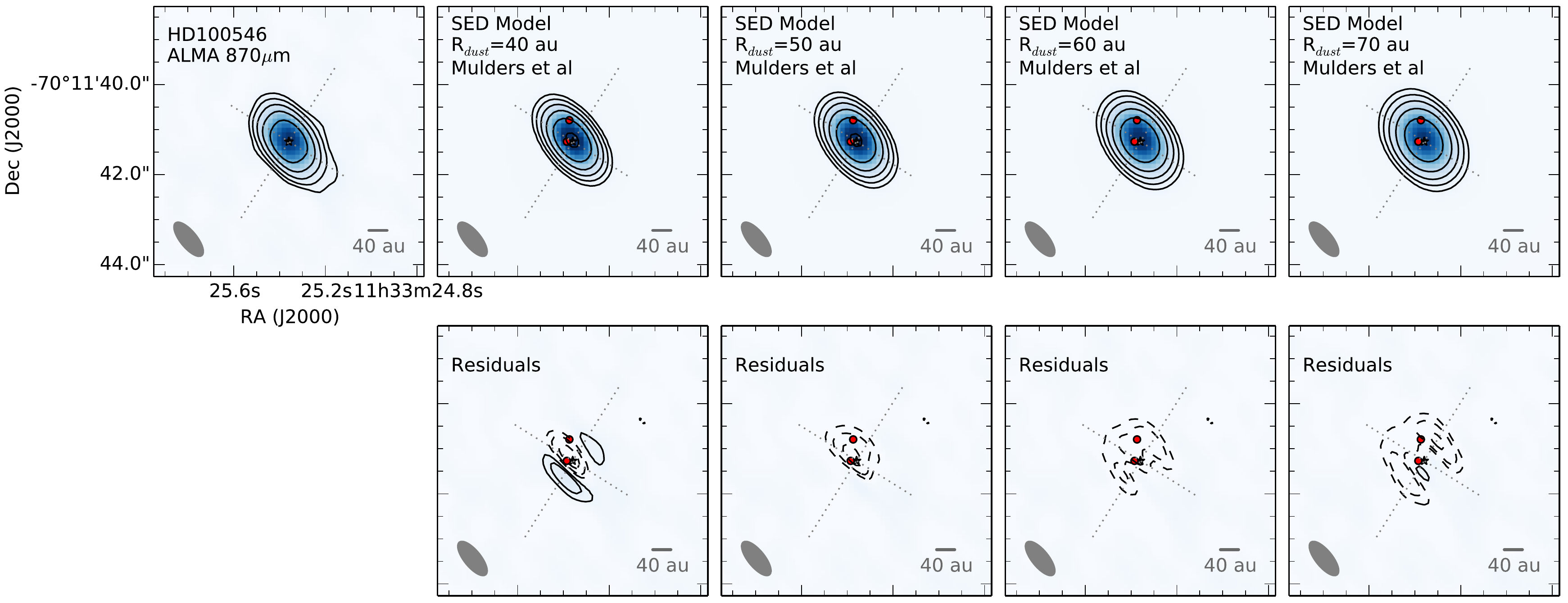}
\caption{\label{fig-cont-data} Top left panel shows the HD\,100546 dust continuum emission 
at 870\,$\mu$m. 
Remaining top row panels show the dust continuum emission predicted by 
models for different outer radii of millimeter-sized dust particles. 
Bottom panels show the difference between data and each model. 
Dotted lines show the major and minor axes direction from the visibility fit.
Contours are shown at [3,6,12,24,48,96]$\times$ rms, where rms is 5.7\,m\jybm, 
with negative contours shown by dashed lines.
Filled circles show the positions of the two planet candidates identified
\citep{brittain2013,Quanz_2013-NACO_HD100546}.
The synthetized beam, 0.93$\arcsec\times$0.37$\arcsec$ (PA=39$\degr$), is shown at bottom left corners.}
\end{figure*}

The 870\,$\mu$m dust continuum emission is compact (see Fig.~\ref{fig-cont-data} top-left panel) 
and dominated by the emission from the outer disk at small radii. 
The total and peak flux are 1.09\,Jy and 0.500\,\jybm, respectively.

We determine the center, major and minor axes, and disk orientation 
by fitting an elliptical Gaussian to the visibility data in CASA 
using the \verb+uvmodelfit+ task. 
The inclination angle is derived from the major and minor axes ratio. 
The results are listed in Table~\ref{table-fit-param}. 
The PA of the major axis (145.14$\degr$$\pm$0.04$\degr$) and inclination angle (41.94$\degr$$\pm$0.03$\degr$), 
are in agreement with those obtained by \cite{ardila2007}, 
who find PA=145$\degr$$\pm$5$\degr$ and i=42$\degr$$\pm$5$\degr$.

\begin{deluxetable}{lc}
\tablewidth{0pt}
\tablecolumns{2}
\tablecaption{Gaussian fit parameters to 870$\mu\textrm{m}$ dust continuum\label{table-fit-param}}
\tablehead{\colhead{Parameter} & \colhead{Value}}

\startdata
R.A. offset\tablenotemark{a} & -398.24$\pm$0.05 (mas)\\
Dec. offset\tablenotemark{a} &  -37.36$\pm$0.05 (mas)\\
Integrated Flux & (1.2015$\pm$0.0002) Jy\\
Major axis (FWHM) & (613.2$\pm$0.1) \,mas\\
Major-to-Minor axis ratio & 0.7439$\pm$0.0004\\
PA & 145.14$\pm$0.04$\degr$\\
i & 41.94$\degr$$\pm$0.03$\degr$
\enddata
\tablenotetext{a}{Position offset measured with respect to pointing phase center: 
(11h33m25.440581s, $-$70d11$\arcmin$41.23633$\arcsec$)
}
\end{deluxetable}

\begin{figure}[h]
\plotone{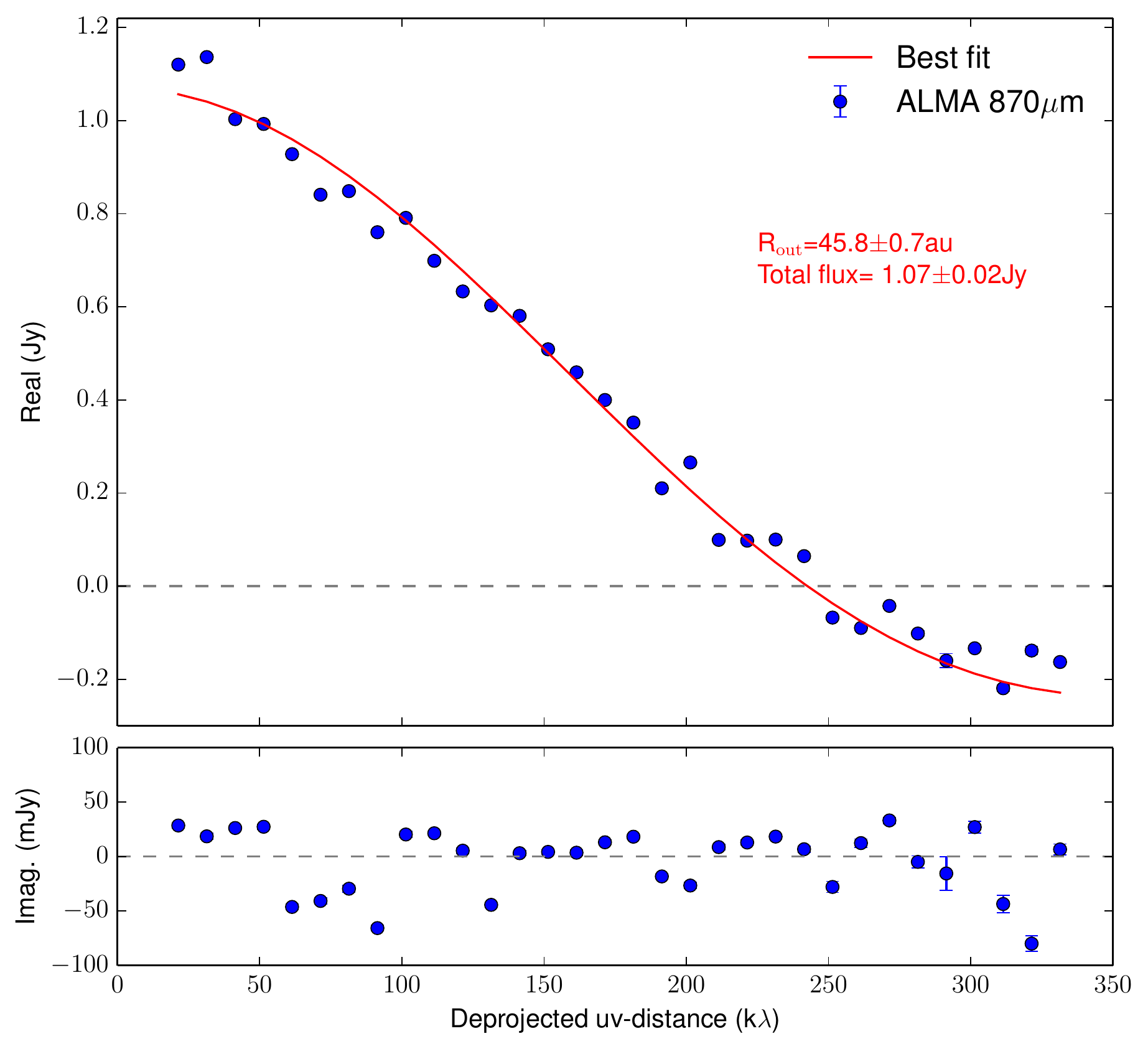}
\caption{\label{fig-cont-uv} De-projected and de-rotated uv-distances. 
Top and bottom panels show the real and imaginary components.
The red curve shows the best fit model for a ring with an inner radius fixed at 14\,au.}
\end{figure}

Fig.~\ref{fig-cont-uv} shows the amplitude as a function of de-rotated and 
de-projected uv-distance. 
The real component (top panel) shows a null at $R_{null}$$\approx$250\,k$\lambda$, 
indicative of a sharp edge in the emission 
\citep[e.g.][]{Andrews_2011-Cavities_Disks, Hughes_2007-TWHydra_Hole}. 
Inspired by the results showing ring-like morphologies in dust emission images 
\citep{Perez_2014-ALMA_Asymmetry,Casassus_2013-HD142527_ALMA,Fukagawa_2013-HD142527_ALMA, vanderMarel_2013-OphIRS48_ALMA, Isella_2013-LkH330}, 
we fit the visibilities with a truncated disk of uniform surface brightness. 
The truncated disk is described by  total flux ($F_0$), inner radius ($R_{\textrm{in}}$), 
and outer radius ($R_{\textrm{out}}$). 
Because of the relatively short baselines probed by these ALMA data it is not possible 
to constrain $R_{\textrm{in}}$, and therefore it is fixed at 14\,au 
\citep[the best fit location of the disk rim as detected in scattered light; Avenhaus et al., submitted; see also][]{quanz2011}. 
The best fit model is shown in Fig.~\ref{fig-cont-uv} by the red curve, 
where the total flux is 1.07$\pm$0.02\,Jy and the outer radius is 45.8$\pm$0.7\,au. 
Clearly, this is an idealized and simple model, and therefore the uncertainties 
for $R_{\textrm{out}}$ are underestimated. 
However, it does reflect the need for the millimeter wavelength emission to be compact 
and with a sharp edge.

To further explore this effect, we compare the observed continuum image with radiative transfer models 
that constrain the dust temperature and structure from SED fitting. 
We use a modified version of the model presented in \cite{Mulders_2011-HD100546_Model}, 
using the radiative transfer and disk modeling code MCMax \citep{Min_2009-MCMax}. 
This model fits the ISO SED up to 200\,$\mu$m and it is consistent 
with Herschel data (SPIRE 200--700\,$\mu$m), 
with a geometry consisting of an inner gap between the star and 0.25\,au, 
an inner accretion disk between 0.25\,au and 1\,au, 
an annular gap from 1 to 13\,au, 
and an outer disk component consisting of micron-sized grains in a 
hydrostatic disk out to $\sim$400\,au, consistent with thermal mid-infared and scattered light imaging. 
To fit the SED at millimeter wavelengths, we add a large grain component (1--3\,mm) to the outer disk 
(starting at 13\,au) with a dust mass of $5\times 10^{-4}$\,M$_\odot$, and 0.1$\times$ 
the hydrostatic scale height, mimicking dust settling. 
To fit the spatial extent of the ALMA image, we truncate the millimeter-sized particles 
at a radius $R_{\textrm{out}}$, 
mimicking effects of radial drift, fragmentation-limited growth or other constraining mechanisms. 
Using a grid of models with $R_{\textrm{out}}$ from 40\,au up to 100\,au and spaced by 10\,au  
the best model is found for 40\,au$<$$R_{\textrm{out}}$$<$60\,au (Figure~\ref{fig-cont-data}).

\subsection{{\rm CO} data}

\begin{figure}[h]
\plotone{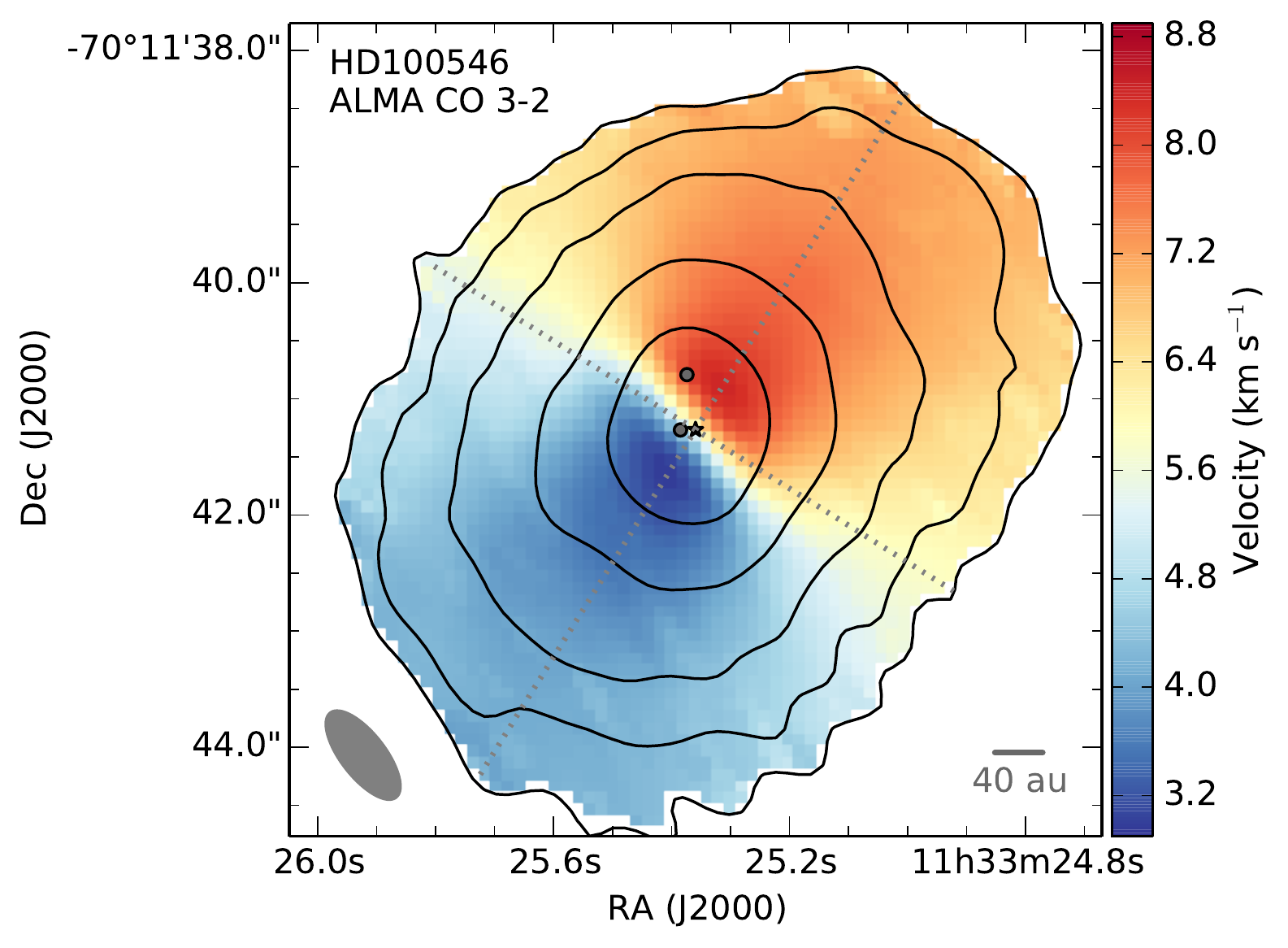}
\caption{The CO~(3--2) moment maps for the HD\,100546 disk. 
The zeroth moment (velocity integrated intensity) map is overlaid in contours shown at 
[3,6,12,24,48,96]$\times$rms, where rms is 0.125\,\jybmkms. 
The first moment (intensity weighted velocity) map is shown in color. 
Dotted lines show the major and minor axes obtained from fitting the dust continuum visibilities. 
Filled circles show the positions of the two planet candidates for HD\,100546 
\citep{brittain2013,Quanz_2013-NACO_HD100546}.
The synthetized beam is shown at the bottom left corner.
\label{fig-CO-data}}
\end{figure}

The CO~(3--2) emission extends out to 3.5$\arcsec$ (340\,au), where the 3-$\sigma$ level is reached, see Fig.~\ref{fig-CO-data}. 
Notice that at this radius the disk temperature, $\approx$30\,K, 
is higher than the CO freeze-out temperature 
\citep[][and references therein]{Dutrey_2014-PPVI_Disks}. 
The disk size from the CO determination is much larger than the size derived 
from the dust emission, but it is consistent with the size required from the 
radiative transfer modeling analysis of single dish CO data 
\citep{Panic_2010-HD100546_Kinematics}.
The zeroth moment (integrated intensity) is calculated between -2.1 and 14.0\,\kms 
and shown by contours in Fig.~\ref{fig-CO-data}. 
The first moment (intensity weighted velocity), shown in Fig.~\ref{fig-CO-data}, 
is calculated using voxels (volume element) with 
brightness $>3\times$rms and in pixels with integrated intensity $>0.4$\,\jybmkms.

\begin{figure}
\plotone{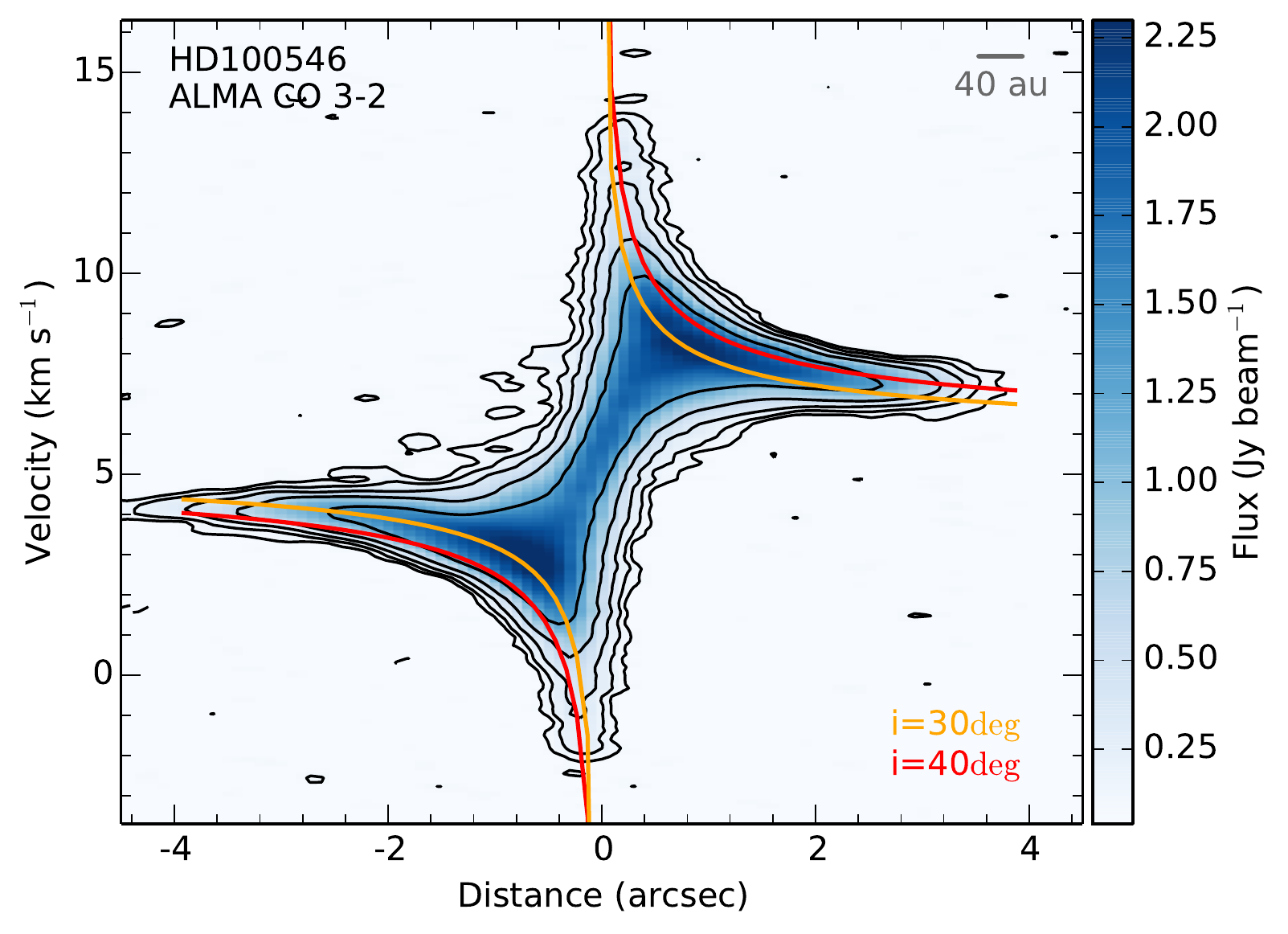}
\caption{PV diagram of CO~(3--2) along the major axis shown in Fig.~\ref{fig-CO-data}. 
Contours are shown at [3,6,12,24,48,96]$\times$rms, 
where rms is 27\,m\jybm per channel. Negative contours are shown by dashed lines. 
Orange and red curves show the expected keplerian velocity for a central star of 2.4\,M$_\odot$ 
and inclination angle of 30$\degr$ and 40$\degr$, respectively.
See  Section~\ref{sec:disk_kinematics} for discussion.
\label{fig-PV-diag}}
\end{figure}

The position velocity (PV) diagram along the disk's major axis is presented in Fig.~\ref{fig-PV-diag}. 
The keplerian velocity profile for the HD\,100546 system, with M$_*$=2.4$\pm$0.1\,M$_\odot$ 
\citep{vandenAncker_1997-HAeBe_Parameters} and 
40$\degr$ inclination angle, reproduces the velocities at a distance $>$2$\arcsec$ from the star 
(red curve in Fig.~\ref{fig-PV-diag}). 
For separations $<$2$\arcsec$ the velocities are better reproduced with an 
inclination angle of 30$\degr$ (orange curve in Fig.~\ref{fig-PV-diag}). 

\section{Discussion}
\subsection{Compact disk of millimeter-sized particles}

From the dust emission model comparison and by the position of the null in the 
visibilities it is clear that the bulk of the millimeter-sized dust particles 
are contained within a 60\,au radius. 
\emph{How is it possible to keep the millimeter-sized dust particles within 60\,au?} 
In order to study how well coupled the millimeter-sized particles are to the gas, 
we calculate the Stokes number, 
$St$$=$$\sqrt{2\pi}\, \rho_s\, a/\Sigma_{gas}$ 
(\citealt{Garaud_2013-Grain_Growth}, see also \citealt{Birnstiel2010_Stokes}), 
where $\rho_s$ is the solid density of particles, 
which we assume to be 1\,g\,cm$^{-3}$ for silicates, $a$$=$1\,mm is the particle size, 
and $\Sigma_{gas}$ is the surface density of the gas disk.  
Using $\Sigma_{gas}(r)$$=$18\,$(r/\textrm{au})$\,g\,cm$^{-2}$ \citep{Mulders_2013-HD100546_Companion_Mass} 
the Stokes number at 50\,au 
is approximately unity, suggesting that millimeter-sized particles are only marginally coupled to the gas.  
Consequently, we expect that they can be subject to radial drift.

Radial drift alone could remove millimeter-sized particles from 
large radii \citep{Weidenschilling__radial_drift,Fouchet_gas_dust_gaps,Birnstiel_2014-Radial_Drift}. 
The net effect of this process is to set an outer truncation radius to the millimeter-sized particles, 
while keeping the small particles and gas disk unaltered. 
Moreover, the suggested presence of a companion in the annular gap ($\sim$1--14\,au) 
would produce a pressure maximum within the outer disk. 
It would enhance particle accumulation and   
reduce radial drift \citep{Pinilla_2012-Ring_Shaped_Disks}. 
This would set an inner truncation radius to the millimeter-sized particle disk. 
Numerical simulations from \cite{Pinilla_2012-Ring_Shaped_Disks} provide an example of a  
truncated disk morphology under the presence of radial drift and an inner planet. 
For their specific star, disk and planet setup 
they find truncated disks (or ring-like structures) of $\sim$20\,au width. 
Future high-resolution observations with ALMA will directly constrain the inner and outer radii 
of the (sub)-millimeter emission.
These data  could also be directly compared to numerical simulations 
appropriate to HD\,100546 and determine if the morphology is consistent with radial drift 
and an inner companion.

\subsection{Interaction with the outer planet candidate}
In addition to radial drift, the presence of the outer protoplanet candidate  \citep{Quanz_2013-NACO_HD100546} 
may also help to clear out the outer disk and generate a sharp outer edge of millimeter-sized particles. 
When a planet interacts with a disk, it will carve out a gap if it is massive enough. 
We hypothesize that a planet may open a gap in the large dust particles, 
since it will form more easily in the millimeter dust than in the gas or smaller dust 
\citep{Fouchet_gas_dust_gaps,Zhu_gas_dust_gaps}.
However, since radial drift already acts to move the millimeter-sized particles inwards on a very fast 
timescale this will effectively mean that the planet acts to open a 'one-sided gap' in 
the millimeter dust. 
In addition, if the planet migrates inwards, this may also help to push the millimeter-sized particles inward and generate the sharper outer edge.
This possible planet-disk interaction scenario can be tested with future high-resolution ALMA observations. 
The outer radius of the dust disk should be the same when observed at different (sub-)millimeter wavelengths (e.g. 3\,mm and 0.45\,mm)  
in the case of a planet-disk interaction (like a one-sided gap opening), while for pure radial drift the outer 
disk radius should be different when observed at different wavelengths.

We further speculate that the presence of both planet candidates may help to constrain the 
dust particles into the apparent ring morphology.

Future observations from ALMA at higher-angular resolution and sensitivity will resolve the innermost disk region 
and yield additional insight into the immediate environment of the disk and perhaps the 
nature of the proposed protoplanet candidates. 
In addition, such observations may help identify the presence and properties of 
potential circumplanetary disks thanks to  
enhanced CO and/or dust continuum emission from the warmer disk material.

\subsection{Disk asymmetries}
The asymmetric model residuals at the 6-$\sigma$ level (bottom panels in Fig.~\ref{fig-cont-data}), 
with the southeastern side being brighter than the northwestern side, 
suggest that the underlying dust continuum emission is asymmetric. 
Asymmetries in millimeter dust emission are not unusual 
\citep{Perez_2014-ALMA_Asymmetry,Casassus_2013-HD142527_ALMA,Fukagawa_2013-HD142527_ALMA, vanderMarel_2013-OphIRS48_ALMA, Isella_2013-LkH330}, 
and they are usually associated with pressure bumps. 
In HD\,100546, there is also an asymmetry between the southeastern and northwestern 
side seen in near-IR scattered light and in mid-IR emission, with a brighter southeastern side 
\citep[Avenhaus et al., submitted;][]{Panic_2014-HD100546_Disk_Asymmetry}. 
Avenhaus et al.  show that the surface brightness profile is steeper in the NW and SW cuts, with a 
change in the power-law index at $\approx$50\,au.
Although only higher-angular resolution observations could confirm it, 
these ALMA data suggest that the dust continuum emission around HD\,100546 is also asymmetric (Fig.~\ref{fig-cont-data}). 
Therefore, any azimuthally symmetric model will not perfectly fit the observations.

\subsection{Disk kinematics}\label{sec:disk_kinematics}	
The morphology seen in the first moment map, Fig.~\ref{fig-CO-data}, where the velocity field in the 
disk inner section is slightly  twisted, 
resembles the velocity maps for warped disks \citep{Rosenfeld_2014-Warped_Disk_Velocity}. 
Although \cite{Rosenfeld_2014-Warped_Disk_Velocity} attempt to reproduce the velocity map for transitional disks 
with radial infall, they show that it is impossible to distinguish between radial infall and a warped disk. 
However, since the twist in the first moment map of HD\,100546 is also observed outside the annular gap,
where infall signatures are not expected, a likely explanation 
is the presence of a warp.

The presence of a warp (with a $\sim$10$\degr$ inclination  change at R$>$200\,au) has previously been suggested 
by \cite{Quillen_2006-HD100546_Warp} and \cite{Panic_2010-HD100546_Kinematics}.
\cite{Quillen_2006-HD100546_Warp} propose that a warp at $\approx$200\,au can explain the 
spiral arm structure seen in scattered light. 
Similarly, a warped disk is suggested by \cite{Panic_2010-HD100546_Kinematics} to explain the 
asymmetric line profiles in single dish observations of different CO transitions and isotopologues.

The position velocity (PV) diagram, Fig.~\ref{fig-PV-diag}, shows that in the outer region (R$>$2$\arcsec$ or 194\,au) 
the inclination angle is consistent with 40$\degr$. 
At smaller distances this inclination predicts too high a velocity, 
and therefore a smaller inclination angle would be preferred. 
However, this same effect is seen in synthetic data of disks with unresolved structure (e.g. spiral arms), 
therefore higher angular resolution observations are need to determine if it is due to unresolved substructure 
or a warped disk.
The inclination angle derived from the 870\,$\mu$m emission (41.94$\degr$) is consistent with the 
inclination angle for the outer radius in the gas disk.

Other possible explanations for the non-keplerian kinematics that might be explored with detailed 
radiative transfer models are: 
a) changes in the flaring angle of the CO disk as a function of radius, which might provide a natural explanation for the 
change in inclination angle of the PV diagram, but that do not necessarily explain the twist in the first moment map; 
b) the presence of spiral arms that are unresolved in these ALMA observations, could produce changes in the 
disk properties (e.g. temperature, radial velocity or disk height) possibly producing variations in 
the first moment map.

\section{Summary and Conclusions}
We have presented the first resolved images of the HD\,100546 transition disk in  thermal dust emission 
and gas as traced by CO~(3--2) using ALMA.
The main results are summarized as follows:
\begin{itemize}
\item The 870$\mu$m dust continuum emission is compact with a radius of $<$60\,au, while the disk emission as traced by 
CO extends up to a radius of $\sim$350\,au. 
The dust emission is well modeled by a truncated disk (ring), where all the emission arises from a region within 
14--60\,au.

\item The dust continuum emission is asymmetric, which is also seen in near-infrared scattered light images. 
This suggests that the underlying dust distribution is not azimuthally symmetric.

\item From the PV diagram an inclination angle of 30$\degr$ and 40$\degr$ are derived for the inner (R$<$200\,au) 
and outer (R$>$200\,au) disk, respectively. However, this effect can also be mimicked by unresolved structure 
(e.g. spirals arms or asymmetries) in the inner disk.

\item We fitted a Gaussian to the dust emission and obtained a position angle of 145.14$\degr$ (E-of-N) and an inclination angle of 41.94$\degr$.
These values are consistent (within the uncertainties) with those from scattered light images.

\item We discuss a possible mechanism to keep millimeter dust particles confined within a 14--60\,au radius. 
Radial drift of millimeter-sized particles will clear the outer disk.
These particles might then be stopped at 14\,au by the presence of a companion in the disk gap.
On the other hand, the presence of the planet companion in the outer disk ($\sim$50--60\,au) may help to clear-out 
the outer disk (similar to a ``one-sided gap''), and it can further help to produce a sharp disk edge. 

\end{itemize}

Future ALMA high-resolution observations can resolve the suggested truncated disk (ring-like) structure and asymmetries. 
These ALMA observations could also provide additional constraints on the existence of the two  protoplanet candidates and 
may even reveal the existence of circumplanetary disks around these young objects.

\acknowledgments

We are grateful to H.M. Schmid, C. Dominik, S. Casassus, and  L. P\'erez for helpful comments. 
We also thank the referee for a thorough and constructive review.
This paper makes use of the following ALMA data: ADS/JAO.ALMA\#2011.0.00863.S. ALMA is a partnership of ESO (representing its member states), NSF (USA) and NINS (Japan), together with NRC (Canada) and NSC and ASIAA (Taiwan), in cooperation with the Republic of Chile. The Joint ALMA Observatory is operated by ESO, AUI/NRAO and NAOJ.
The National Radio Astronomy Observatory is a facility of the National Science Foundation operated under cooperative agreement by Associated Universities, Inc.
JEP is supported by the Swiss National Science Foundation, project number CRSII2\_141880.
FM is supported by the ETH Zurich Postdoctoral Fellowship Program and the 
Marie Curie Actions for People COFUND program.
OP is supported by the European Union through ERC grant number 279973.
This research made use of Astropy, a community-developed core Python package for Astronomy \citep{Astropy_2013}, 
and APLpy, an open-source plotting package for Python hosted at \url{http://aplpy.github.com}.

Facilities: \facility{ALMA}

\end{document}